\documentclass[aps,pre,showpacs,twocolumn,floatfix]{revtex4}
\usepackage{graphicx}
\usepackage{amssymb}

\usepackage{wasysym}

\begin{document}

\title{Tunable heat pump by modulating the coupling to the leads}
\author{Eduardo C. Cuansing}
\affiliation{Department of Electrical and Computer Engineering, National
             University of Singapore, Singapore 117576, Republic of
             Singapore}
\author{Jian-Sheng Wang}
\affiliation{Department of Physics and Centre for Computational Science
             and Engineering, National University of Singapore,
             Singapore 117542, Republic of Singapore}

\date{27 May 2010}

\begin{abstract}

We follow the nonequilibrium Green's function formalism to study
time-dependent thermal transport in a linear chain system consisting of 
two semi-infinite leads connected together by a coupling that is 
harmonically modulated in time. The modulation is driven by an external 
agent that can absorb and emit energy. We determine the energy current 
flowing out of the leads exactly by solving numerically the Dyson 
equation for the contour-ordered Green's function. The amplitude of the
modulated coupling is of the same order as the interparticle coupling
within each lead. When the leads have the same temperature, our numerical
results show that modulating the coupling between the leads may direct 
energy to either flow into the leads simultaneously or flow out of the 
leads simultaneously, depending on the values of the driving frequency 
and temperature. A special combination of values of the driving frequency 
and temperature exists wherein no net energy flows into or out of the 
leads, even for long times. When one of the leads is warmer than the 
other, net energy flows out of the warmer lead. For the cooler lead, 
however, the direction of the energy current flow depends on the values
of the driving frequency and temperature. In addition, we find transient 
effects to become more pronounced for higher values of the driving 
frequency.

\end{abstract}


\maketitle

\section{Introduction}
\label{sec:intro}

The transport of phonons in mesoscale and nanoscale devices is an
important issue relevant to the questions of heat generation in 
devices and their structural stability. Experiments measuring the 
heat generated in electric current-carrying metal-molecule junctions 
found that the generated heat can be substantial \cite{junctions} and 
can therefore threaten the device's integrity. Efficiently dissipating 
heat in such devices is thus important and a problem that must be 
considered. A way of manipulating heat in nanoscale devices is by 
utilizing a heat pump that directs heat from one part of the device to 
another or to an external reservoir by means of an applied external 
work. Models on the mechanism of such a nanoscale heat pump have been 
proposed in systems where the pump works against the thermal gradient 
between two reservoirs in the system \cite{gradients,ai10,forcedriven} 
and in systems where there is no net thermal bias between the two 
reservoirs \cite{nakagawa06,nobias,ren10}. In addition, other models 
employing quantum particle pumps that differentiate and filter hot and 
cold particles have been proposed \cite{quantumpumps}. In this paper we 
present an alternative model of a phonon pump that directs energy, and 
thus heat, into or out of regions of the device by dynamically modulating 
the coupling between those regions. The model is different from previous 
models of heat pumps where requirements of either modulating the
temperatures of reservoirs \cite{nobias,ren10}, or having an external
driving force acting at the central portion of the device
\cite{ai10,forcedriven,nakagawa06}, or filtering particles according
to their temperatures \cite{gradients,quantumpumps} should be
satisfied. Our model, in comparison, requires an external agent that can 
either absorb or release heat and is dynamically modulating the coupling 
between two parts of the device. A thermal gradient is not necessary for 
our model phonon pump to work.

To induce a phonon pump action in our model the coupling between 
the two parts of the system is harmonically modulated in time. 
Experimentally, this can be done by, for example, harmonically varying
the distance between two molecules, therefore modulating the
coupling between the molecules. Time-dependent transport of phonons in 
molecular systems, however, is a topic that is not yet fully understood
theoretically. Although our understanding of the subject has improved 
tremendously during the past few years, most of the results pertain to 
steady-state and long-time behavior \cite{wang08}. Time-dependent phonon 
transport with non-adiabatic and strong perturbations has recently been 
studied in a thermal switch device where the coupling to the reservoirs 
is abruptly turned on \cite{cuansing10}. In this paper we extend the 
thermal switch model to one where the reservoir coupling is modulated 
in time. The system subsequently acts as a phonon pump that can be tuned 
by varying the frequency of modulation of the coupling to the reservoirs.

\begin{figure}[h!]
\includegraphics[width=3.2in]{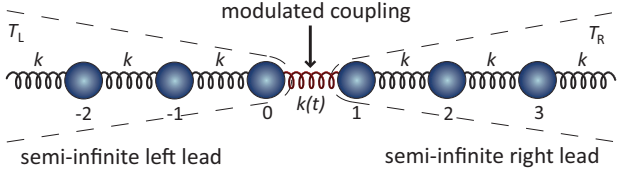}
\caption{(Color online) An illustration of an infinite linear chain 
whose two semi-infinite parts are connected by a coupling $k(t)$ that is 
modulated in time. The labels of the first three particles in each lead 
are shown beside each particle. Nearest-neighbor particles within the
leads interact via an interparticle spring constant $k$. An on-site
spring with spring constant $k_0$ is also experienced by each particle.
\label{fig:chain}}
\end{figure}

\section{Model and theoretical approach}
\label{sec:theory}

In this paper we consider phonon transport in a one-dimensional chain. 
Shown in Fig.~\ref{fig:chain} is a linear chain consisting of two
semi-infinite leads, or reservoirs, coupled together by a coupling
$k(t)$ that is modulated in time. The particles in each lead are
coupled to their nearest neighbors by a coupling constant $k$. In
addition, each particle experiences an on-site potential with spring 
constant $k_0$. The temperatures in the left and right leads are 
$T_{\rm L}$ and $T_{\rm R}$, respectively. The particles can vibrate only 
along the horizontal axis. Particles in each lead follow a Hamiltonian 
with purely harmonic, nearest-neighbor, interactions that do not vary in 
time:
\begin{equation}
H^{\alpha} = \frac{1}{2} \sum_i \left(\dot{u}_i^{\alpha}\right)^2
  + \frac{1}{2} \sum_{ij} u_i^{\alpha} K_{ij}^{\alpha} u_j^{\alpha},~~~
  \alpha = {\rm L,R},
\label{eq:leads}
\end{equation}
where the first sum is over all particles and the second sum is over 
all nearest-neighbor pairs in the leads. Each nearest-neighbor pair,
however, is considered twice and so we divide the second term by two. The 
transformed coordinates $u_i = \sqrt{m} x_i$ is used, where $x_i$ is the 
relative displacement of the $i$-th particle of mass $m$, and 
${\bf K}^{\alpha}$ is the coupling matrix. This matrix is from the 
dynamic matrix of the full system:
\begin{equation}
{\bf K} = \left( \begin{array}{cc}
           {\bf K}^{\rm L} & {\bf V}^{\rm LR} \\
           {\bf V}^{\rm RL} & {\bf K}^{\rm R}
           \end{array} \right),
\label{eq:dynamic}
\end{equation}
where in the one-dimensional chain the spring constant matrices 
${\bf K}^{\rm L}$ and ${\bf K}^{\rm R}$ are semi-infinite tridiagonal 
sub-matrices consisting of $2 k + k_0$ along the diagonal and $-k$ along 
both off-diagonals. The coupling matrices ${\bf V}^{\rm LR}(t)$ and 
${\bf V}^{\rm RL}(t)$ are the couplings to the leads that vary in time. 
The Hamiltonian involving the lead coupling is:
\begin{equation}
H^{\rm LR}(t) = \sum_{ij} u_i^{\rm L}\, V_{ij}^{\rm LR}(t)\, u_j^{\rm R},
\label{eq:coupling}
\end{equation}
where the sum is over all particles in each lead that are directly
coupled to the other lead. For the linear chain shown in 
Fig.~\ref{fig:chain} the coupling matrices each have only one nonzero 
element:
\begin{equation}
V_{01}^{\rm LR}(t) = -k(t)~~{\rm and}~~V_{10}^{\rm RL}(t) = -k(t).
\label{eq:chainpotential}
\end{equation} 
Notice that during the same instant in time, 
$H^{\rm LR}(t) = H^{\rm RL}(t)$. The total time-dependent Hamiltonian 
for the two-lead system is 
\begin{equation}
H(t) = H^{\rm L} + H^{\rm R} + H^{\rm LR}(t).
\label{eq:sumhamiltonian}
\end{equation}
Note that what we have is an open system consisting of the two leads 
and their coupling that is being modulated by an external agent. The
energy, therefore, that this two-lead system gains or loses is coming 
from or going to the external agent.

The energy current flowing out of the left lead is
\begin{equation}
I^{\rm L}(t) = -\left< \frac{dH^{\rm L}}{dt} \right> = \frac{i}{\hbar}
  \left< \left[ H^{\rm L},H \right] \right>,
\label{eq:current1}
\end{equation}
i.e., it is the negative of the expectation value of the rate of change 
in $H^{\rm L}$. The Heisenberg equation of motion is used in the second
equality. The position and momentum of a particle obey the commutation 
relation
\begin{equation}
\left[ u_i^{\alpha}(t), \dot{u}_j^{\beta}(t) \right] = i \hbar\,\delta_{ij}
  \,\delta^{{\alpha}{\beta}},~~~~~\alpha, \beta = {\rm L, R},
\label{eq:commutation}
\end{equation}
where $i$ and $j$ are particle labels. Note that this commutation 
relation only exists at the same instant in time. We thus find that 
the only term in $H$ that does not commute with $H^{\rm L}$ is $H^{\rm LR}$.
Equation~(\ref{eq:current1}) therefore becomes
\begin{equation}
I^{\rm L}(t) = \frac{i}{2\hbar} \sum_{jmn}\left< \left[ \dot{u}_j^{\rm L} 
  \,\dot{u}_j^{\rm L}, u_m^{\rm L}\,V_{mn}^{\rm LR}\,u_n^{\rm R} \right] 
  \right>,
\label{eq:sametime}
\end{equation}
where all the terms in the right-hand side occur at the same time $t$. 
Now define the real-time lesser Green's function as
\begin{equation}
G_{ij}^{{\rm RL},<}(t_1,t_2) = -\frac{i}{\hbar}\left< u_j^{\rm L}(t_2)
  \,u_i^{\rm R}(t_1) \right>.
\label{eq:lesser}
\end{equation}
We would like to use this Green's function to properly calculate the 
current. Re-expressing Eq.~(\ref{eq:sametime}) using two time variables
we get
\begin{eqnarray}
I^{\rm L}(t) & = & \frac{i}{2\hbar} \sum_{jmn}\,\left< \left[ 
  \dot{u}_j^{\rm L}(t_2)\,\dot{u}_j^{\rm L}(t_2),\right.\right. \nonumber \\
  & & \left.\left.\left.u_m^{\rm L}(t_1)\,V_{mn}^{\rm LR}(t_1)
  \,u_n^{\rm R}(t_1) \right] \right> \right|_{t_1 = t_2 = t},
\label{eq:changetime}
\end{eqnarray}
where we set $t_1 = t_2 = t$ in the end. Since 
$V^{\rm LR}(t) = V^{\rm RL}(t)$, we have
\begin{eqnarray}
I^{\rm L}(t) & = & \frac{i}{2 \hbar}\,\sum_{jmn}  
  \,c_{jm}(t_2,t_1)\,V_{mn}^{\rm LR}(t_1) \,\frac{\partial}{\partial t_2}
  \left\{\left< u_j^{\rm L}(t_2)\,u_n^{\rm R}(t_1) \right>\right.
  \nonumber \\
  & & +\left.\left.\left< u_n^{\rm R}(t_1)\,u_j^{\rm L}(t_2) \right> 
  \right\} \right|_{t_1 = t_2 = t},
\label{eq:nogreen}
\end{eqnarray}
where 
$c_{jm}(t_2,t_1) = \left[\dot{u}_j^{\rm L}(t_2), u_m^{\rm L}(t_1)\right]$.
Making use of the lesser Green's function, Eq.~(\ref{eq:lesser}), and its 
complex conjugate, we get
\begin{eqnarray}
I^{\rm L}(t) & = & -i \sum_{jmn}\,c_{jm}(t_2,t_1)\,V_{mn}^{\rm LR}(t_1) 
  \nonumber \\
  & & \times\left.{\rm Im}\!\left\{\frac{\partial}{\partial t_2} 
  G_{nj}^{{\rm RL},<}(t_1,t_2) \right\}\right|_{t_1 = t_2 = t},
\label{eq:generalcurrent}
\end{eqnarray}
where ``Im'' means taking the imaginary part. 
Equation~(\ref{eq:generalcurrent}) is a general equation for the energy 
current flowing out of the left lead. Note that the result is independent 
of the dimension of the system. For the one-dimensional linear chain setup 
shown in Fig.~\ref{fig:chain} we use Eq.~(\ref{eq:chainpotential}) and
the fact that $c_{jm}(t,t) = -i \hbar\,\delta_{jm}$ to get
\begin{equation}
I^{\rm L}(t) = \hbar\left.k(t)\,{\rm Im}\!\left\{\frac{\partial}
  {\partial t_2}G_{10}^{{\rm RL},<}(t_1,t_2) \right\} 
  \right|_{t_1 = t_2 = t}.
\label{eq:leftcurrent}
\end{equation}
Equation~(\ref{eq:leftcurrent}) is the primary equation we use to calculate 
the current flowing out of the left lead. Similarly, we define the energy
current flowing out of the right lead as 
$I^{\rm R}(t) = -\left< dH^{\rm R}/dt \right>$ and derive an
expression in terms of the lesser Green's function:
\begin{equation}
I^{\rm R}(t) = \hbar\left.k(t)\,{\rm Im}\!\left\{\frac{\partial}
  {\partial t_2}G_{01}^{{\rm LR},<}(t_1,t_2) \right\} 
  \right|_{t_1 = t_2 = t}.
\label{eq:rightcurrent}
\end{equation}

To calculate the currents in Eqs.~(\ref{eq:leftcurrent}) and
(\ref{eq:rightcurrent}) we need to determine the nonequilibrium lesser
Green's functions in those equations. We now follow the Schwinger-Keldysh 
formalism, in which a complex-time contour is employed 
\cite{schwinger-keldysh,rammer86,jauho94,haug08}, to determine the Green's 
functions.

\begin{figure}[h!]
\includegraphics[width=1.4in]{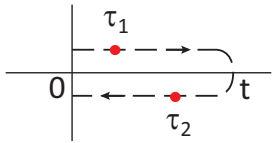}
\caption{(Color online) The complex-time contour in the Keldysh formalism. 
The path of the contour begins at time $t = 0$, goes to time $t$, and 
then goes back to time $t = 0$. $\tau_1$ and $\tau_2$ are complex-time 
variables along the contour.
\label{fig:contour}}
\end{figure}

Shown in Fig.~\ref{fig:contour} is the Keldysh contour we use. At times 
$t < 0$ we consider the left and right leads to be decoupled and each 
is in thermal equilibrium with temperature $T_{\rm L}$ and $T_{\rm R}$, 
respectively. At time $t = 0$ the coupling between the leads, $k(t)$, 
is switched on. We then calculate the energy current at time $t$, i.e., 
the time at the right edge of the contour. Note that instead of adiabatic
switch-on, the coupling $k(t)$ is abruptly turned on at $t = 0$. In 
addition, the left and right leads are uncorrelated before $t = 0$ and 
so there is no imaginary tail when the contour goes back to time $t = 0$. 
We now define the contour-ordered Green's function
\begin{equation}
G_{ij}^{\rm RL}(\tau_1,\tau_2) = -\frac{i}{\hbar}\left<{\rm T}_c 
  \,u_i^{\rm R}(\tau_1)\,u_j^{\rm L}(\tau_2) \right>,
\label{eq:contourgreen}
\end{equation}
where the ${\rm T}_c$ is the contour-ordering operator, $\tau_1$ and 
$\tau_2$ are contour variables, and $u_i^{\rm R}$ and $u_j^{\rm L}$ are
operators in the Heisenberg picture. Transforming to the interaction 
picture, we separate the Hamiltonian in Eq.~(\ref{eq:sumhamiltonian}) 
into the free-particle quadratic part, $H_0 = H^{\rm L} + H^{\rm R}$, and 
the interaction part, $H_{\rm int}(t) = H^{\rm LR}(t)$. We then write
the contour-ordered Green's function in the interaction picture as
\begin{equation}
G_{ij}^{\rm RL}(\tau_1,\tau_2) = -\frac{i}{\hbar}\left<{\rm T}_c
  \,e^{-i/\hbar\,\int_c H_{\rm int}(\tau') d\tau'}\,u_i^{\rm R}(\tau_1)
  \,u_j^{\rm L}(\tau_2)\right>,
\label{eq:interaction}
\end{equation}
where $c$ is the contour shown in Fig.~\ref{fig:contour} and the
average is now taken with respect to the equilibrium distributions
when $t < 0$. We expand the exponential to perform a perturbation
expansion. For the $0$th-order term we find
\begin{equation}
G_{ij,0}^{\rm RL}(\tau_1,\tau_2) = -\frac{i}{\hbar}\left<{\rm T}_c
  u_i^{\rm R}(\tau_1)\,u_j^{\rm L}(\tau_2)\right> = 0
\label{eq:0th}
\end{equation}
because there is no coupling term that would connect the left and 
right particles. The $1$st-order term is
\begin{widetext}
\begin{eqnarray}
G_{ij,1}^{\rm RL}(\tau_1,\tau_2) & = & \left(-\frac{i}{\hbar}\right)^2
  \sum_{mn}\int_c d\tau'\,V_{mn}^{\rm LR}(\tau') \left<{\rm T}_c
  u_m^{\rm L}(\tau')\,u_n^{\rm R}(\tau')\,u_i^{\rm R}(\tau_1)\,
  u_j^{\rm L}(\tau_2)\right>, \nonumber \\
  & = & \sum_{mn}\int_c d\tau'\,\left\{-\frac{i}{\hbar}\left<{\rm T}_c
  u_m^{\rm L}(\tau')\,u_j^{\rm L}(\tau_2)\right>\right\} 
  V_{mn}^{\rm LR}(\tau') \left\{-\frac{i}{\hbar}\left<{\rm T}_c 
  u_n^{\rm R}(\tau')\,u_i^{\rm R}(\tau_1)\right> \right\},
\label{eq:1st}
\end{eqnarray} 
\end{widetext}
where Wick's theorem is used to expand the four-particle average into
two two-particle averages. There are actually three different ways to
expand the four-particle average but the other two configurations 
vanish and we are left only with the term shown in the second equality. 
We would like to note that the use of Wick's theorem here is justified 
because the expansion is with respect to the quadratic $H_0$. 

Define the equilibrium Green's function of the free leads as
\begin{equation}
g_{ij}^{\alpha}(\tau_1,\tau_2) = -\frac{i}{\hbar}\left<{\rm T}_c
  u_i^{\alpha}(\tau_1)\,u_j^{\alpha}(\tau_2)\right>,~~~\alpha = 
  {\rm L, R},
\label{eq:equil}
\end{equation}
where the average is taken with respect to $H_0$, i.e., the 
corresponding equilibrium distribution of the leads. Note that unlike 
the nonequilibrium $G^{\rm RL}$, the equilibrium $g^{\alpha}$ satisfies 
time-translation invariance and thus its Fourier transform exists and 
can be calculated. Writing Eq.~(\ref{eq:1st}) in terms of the 
equilibrium Green's functions we get
\begin{equation}
G_{ij,1}^{\rm RL}(\tau_1,\tau_2) = \sum_{mn} \int_c d\tau'
  g_{in}^{\rm R}(\tau_1,\tau')\,V_{nm}^{\rm RL}(\tau')\,
  g_{mj}^{\rm L}(\tau',\tau_2).
\label{eq:1stequil}
\end{equation}
To get the lesser version of the nonequilibrium $G_{ij,1}^{\rm RL}$ we 
employ analytic continuation and Langreth's theorem \cite{haug08} to get
\begin{eqnarray}
G_{ij,1}^{{\rm RL},<}(t_1,t_2) & = & \sum_{mn} \int_0^t dt'
  \,V_{nm}^{\rm RL}(t') \nonumber \\ 
  & & \times~\left\{g_{in}^{{\rm R},r}(t_1,t')\,g_{mj}^{{\rm L},<}(t',t_2)
  \right. \nonumber \\
  & & +~\left.g_{in}^{{\rm R},<}(t_1,t')\,g_{mj}^{{\rm L},a}(t',t_2)\right\},
\label{eq:langreth}
\end{eqnarray}
where $g_{in}^{{\rm R},r}$ and $g_{mj}^{{\rm L},a}$ are the retarded and 
advanced versions of the equilibrium Green's functions, respectively. 
Similarly, the retarded and advanced versions of the first-order 
nonequilibrium Green's functions are
\begin{equation}
G_{ij,1}^{{\rm RL},\zeta} = \sum_{mn} \int_0^t dt'
  \,g_{in}^{{\rm R},\zeta}(t_1,t')\,V_{nm}^{\rm RL}(t')
  \,g_{mj}^{{\rm L},\zeta}(t',t_2),
\label{eq:retadv}
\end{equation}
where $\zeta = r,a$. For the one-dimensional chain shown in 
Fig.~\ref{fig:chain} only the $i = 1$ and $j = 0$ label combination is 
nonzero. In addition, the coupling potential in 
Eq.~(\ref{eq:chainpotential}) is nonzero only when $n = 1$ and
$m = 0$ in the sum. All the other combinations of the indices do not
contribute.

\begin{figure}[h!]
\includegraphics[width=2.5in]{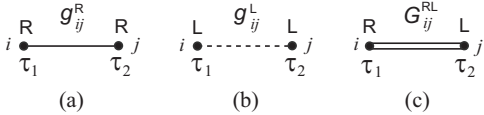}
\caption{Diagram representations for the equilibrium Green's functions 
(a) $g_{ij}^{\rm R}(\tau_1,\tau_2)$ and (b) $g_{ij}^{\rm L}(\tau_1,\tau_2)$, 
and (c) the nonequilibrium Green's function $G_{ij}^{\rm RL}(\tau_1,\tau_2)$.
\label{fig:diagrams}}
\end{figure}

\begin{figure}[h!]
\includegraphics[width=3.2in]{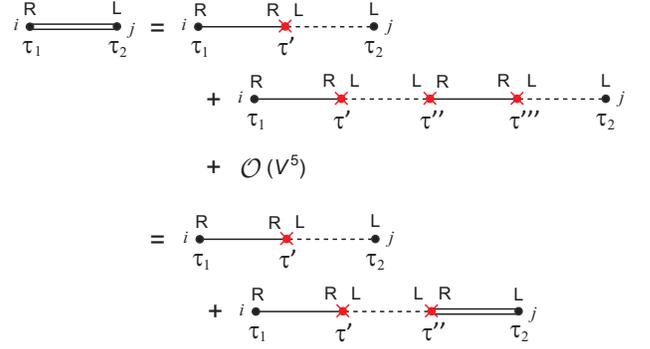}
\caption{(Color online) Diagrammatic perturbation expansion of $G^{\rm RL}$ 
in terms of $g^{\rm R}$ and $g^{\rm L}$. Each interaction vertex, i.e., the 
crossed-out (red) dot, includes an interaction potential $V^{\rm RL}$ 
and an integration with respect to the internal complex-time variable. The 
second equality is the diagram representation of the Dyson equation.
\label{fig:expansion}}
\end{figure}

To facilitate the calculation of higher-order terms in the perturbation 
expansion we utilize a diagrammatic approach. Shown in 
Fig.~\ref{fig:diagrams} are the diagrams representing the relevant 
contour-ordered Green's functions in the expansion. A diagram can not be 
constructed for the zeroth-order term, Eq.~(\ref{eq:0th}), since it 
contains a $u^{\rm L}$ and a $u^{\rm R}$ pair. The first-order term, in
contrast, has just the right number of double pairs of $u^{\rm L}$ and 
$u^{\rm R}$ and a $V^{\rm RL}$ to connect the two equilibrium Green's 
functions. The second-order term has the same shortcoming as the 
zeroth-order term, i.e., there is an extra $u^{\rm L}$ and $u^{\rm R}$ 
pair. In fact, all the rest of the even-ordered terms have the same 
extra $u^{\rm L}$ and $u^{\rm R}$ pair. All of the even-ordered terms 
therefore do not contribute to the perturbation expansion. As for the 
odd-ordered terms, they consist of repetitions of the first-order 
diagram. Shown in Fig.~\ref{fig:expansion} is the perturbation expansion 
in diagram representation.

In the first equality of Fig.~\ref{fig:expansion}, the first and 
third-order terms are explicitly shown. Notice that the third-order term 
consists of two first-order terms. The diagram representation of the Dyson 
equation is shown in the second equality of Fig.~\ref{fig:expansion}. Note 
that this equation includes all terms in the expansion. Based on the 
diagrammatic rules, we can now write the Dyson equation for the 
nonequilibrium Green's function as
\begin{eqnarray}
\lefteqn{G_{ij}^{\rm RL}(\tau_1,\tau_2) = \sum_{mn} \int_c d\tau'
  \,g_{im}^{\rm R}(\tau_1,\tau')\,V_{mn}^{\rm RL}(\tau')
  \,g_{nj}^{\rm L}(\tau',\tau_2)} \nonumber \\
  & & +~\sum_{mnpq} \int_c d\tau' \int_c d\tau''
  \,g_{im}^{\rm R}(\tau_1,\tau')\,V_{mn}^{\rm RL}(\tau')
  \,g_{np}^{\rm L}(\tau',\tau'') \nonumber \\
  & & ~~~\times~V_{pq}^{\rm LR}(\tau'')\,G_{qj}^{\rm RL}(\tau'',\tau_2).
\label{eq:dyson}
\end{eqnarray}
Applying Langreth's theorem \cite{haug08} to Eq.~(\ref{eq:dyson}) and 
then iterating the resulting equation, we get a closed-form formula that 
includes all orders of the expansion:
\begin{eqnarray}
\lefteqn{G_{ij}^{{\rm RL},<}(t_1,t_2) = G_{ij,1}^{{\rm RL},<}(t_1,t_2)} 
  \nonumber \\
  & & +~\sum_{mn} \int_0^t dt'\,G_{im}^{{\rm RL},r}(t_1,t')
  \,V_{mn}^{\rm LR}(t')\,G_{nj,1}^{{\rm RL},<}(t',t_2) \nonumber \\
  & & +~\sum_{mn} \int_0^t dt'\,G_{im,1}^{{\rm RL},<}(t_1,t')
  \,V_{mn}^{\rm LR}(t')\,G_{nj}^{{\rm RL},a}(t',t_2) \nonumber \\
  & & +~\sum_{mnpq} \int_0^t dt'\,\int_0^t dt''\,
  G_{im}^{{\rm RL},r}(t_1,t')\,V_{mn}^{\rm LR}(t') \nonumber \\
  & & ~~~\times~G_{np,1}^{{\rm RL},<}(t',t'')\,V_{pq}^{\rm LR}(t'')
  \,G_{qj}^{{\rm RL},a}(t'',t_2).
\label{eq:green}
\end{eqnarray}
Equation~(\ref{eq:green}) is the exact formula for the general 
nonequilibrium Green's function needed to calculate the current flowing 
through the left lead in Eq.~(\ref{eq:generalcurrent}). For the linear 
chain shown in Fig.~\ref{fig:chain}, we use the coupling potential in
Eq.~(\ref{eq:chainpotential}) and the indices $i = n = q = 1$ and
$j = m = p = 0$.

To solve Eq.~(\ref{eq:green}) we need to determine 
$G_{ij,1}^{{\rm RL},<}$, $G_{ij}^{{\rm RL},a}$, and 
$G_{ij}^{{\rm RL},r}$. From Eq.~(\ref{eq:langreth}) the first-order
nonequilibrium Green's function can be calculated by
\begin{eqnarray}
G_{10,1}^{{\rm RL},<}(t_1,t_2) & = & -\int_0^t dt'\,k(t')\,
  \left\{g_{11}^{{\rm R},r}(t_1,t')\,g_{00}^{{\rm L},<}(t',t_2) 
  \right.\nonumber \\
  & & \left.+~g_{11}^{{\rm R},<}(t_1,t')\,g_{00}^{{\rm L},a}(t',t_2) 
  \right\}.
\label{eq:lessernonequil}
\end{eqnarray}
To determine the full retarded Green's function we apply Langreth's
theorem to Eq.~(\ref{eq:dyson}) to get the equation
\begin{eqnarray}
G_{10}^{{\rm RL},r}(t_1,t_2) & + & \int_0^t dt'\,k(t')
  \,G_{10,1}^{{\rm RL},r}(t_1,t')\,G_{10}^{{\rm RL},r}(t',t_2)
  \nonumber \\
  & = & G_{10,1}^{{\rm RL},r}(t_1,t_2),
\label{eq:retnonequil}
\end{eqnarray}
where Eq.~(\ref{eq:retadv}) is used for the first-order Green's
functions. A similar equation can be derived for the full advanced 
Green's function. These two equations for the full retarded and 
advanced Green's functions can be solved by the process discussed in 
Sec.~\ref{sec:numerics}.

\section{Numerically calculating the energy current}
\label{sec:numerics}

In the linear chain, the energy flowing out of the left lead, 
$I^{\rm L}(t)$, at time $t$ can be calculated using 
Eq.~(\ref{eq:leftcurrent}), the nonequilibrum lesser Green's function 
$G^{{\rm RL},<}(t_1,t_2)$ shown in Eq.~(\ref{eq:green}), and its 
derivative with respect to $t_2$. From Eq.~(\ref{eq:green}), the 
$G^{{\rm RL},<}$ can be calculated from the first-order nonequilibrium
lesser Green's function $G_1^{{\rm RL},<}$, the full nonequilibrium
retarded and advanced Green's functions, $G^{{\rm RL},r}$ and 
$G^{{\rm RL},a}$, respectively, and their derivatives with respect to 
$t_2$. Furthermore, from Eq.~(\ref{eq:langreth}), the $G_1^{{\rm RL},<}$ 
can be calculated from the integral of equilibrium Green's functions 
$g^{{\rm R},r}$, $g^{{\rm L},<}$, $g^{{\rm R},<}$, and $g^{{\rm L},a}$. 
In addition, from Eq.~(\ref{eq:retnonequil}), the full nonequilibrium 
retarded and advanced Green's functions can be calculated from the 
integral of equilibrium Green's functions. All of the nonequilibrium 
Green's functions, therefore, can be calculated from the integrals of 
equilibrium Green's functions. What we ultimately need then are the 
equilibrium Green's functions.

Equilibrium Green's functions satisfy time-translation invariance
and therefore their Fourier transforms exist. In frequency space the 
retarded equilibrium Green's functions for the semi-infinite linear 
chain leads are known to be \cite{wang07}
\begin{equation}
g_{ij}^{{\alpha},r}[\omega] = -\frac{\lambda}{k}\,\lambda^{|i-j|},
  ~~~\alpha = {\rm L, R},
\label{eq:retequil}
\end{equation}
where the square brackets mean that the function is a Fourier transform
and
\begin{equation}
\lambda = -\frac{\Omega}{2 k} \pm \frac{1}{2 k} \sqrt{\Omega^2 - 4 k^2},
\label{eq:lambda}
\end{equation}
where $\Omega = (\omega + i \eta)^2 - 2 k - k_0$ and the choice
between the plus or minus sign depends on satisfying $|\lambda| < 1$. 
Furthermore, the lesser equilibrium Green's function can be determined 
from \cite{wang07}
\begin{equation}
g_{ij}^{{\alpha},<}[\omega] = 2 i f_{\alpha}\,{\rm Im}\!\left\{
  g_{ij}^{{\alpha},r}[\omega]\right\},
\label{eq:lessequil}
\end{equation}
where $f_{\alpha}$ is the Bose-Einstein distribution function of the 
$\alpha$ lead. Given a function $F[\omega]$ in frequency space its
inverse Fourier transform is
\begin{equation}
F(t_1,t_2) = \int_{-\infty}^{\infty} \frac{d\omega}{2 \pi}\,F[\omega]
  \,e^{-i \omega (t_1 - t_2)}.
\label{eq:invFT}
\end{equation}
We can thus take the inverse Fourier transform of 
Eqs.~(\ref{eq:retequil}) and (\ref{eq:lessequil}) to determine the 
time-dependence of the corresponding equilibrium Green's functions.
In addition, the advanced equilibrium Green's functions can be
calculated from the retarded version by \cite{wang08}
\begin{equation}
g_{ij}^{{\alpha},a}[\omega] = \left(g_{ji}^{{\alpha},r}[\omega]
  \right)^{\ast}.
\label{eq:advequil}
\end{equation}
The integrals appearing in the inverse Fourier transforms are
numerically calculated using the trapezoidal rule \cite{press07}.

After numerically calculating the equilibrium Green's functions, we
can use Eq.~(\ref{eq:lessernonequil}) to determine the first-order
nonequilibrium lesser Green's function $G_{ij,1}^{{\rm RL},<}$. We again 
use the trapezoidal rule to calculate the integral in 
Eq.~(\ref{eq:lessernonequil}).

To calculate the full nonequilibrium retarded Green's function 
$G_{ij}^{{\rm RL},r}$, we solve Eq.~(\ref{eq:retnonequil}). This 
equation is in the form of a Fredholm equation of the second kind
\cite{press07}
\begin{equation}
f(t_a,t_b) + \int_0^t dt'\,f_1(t_a,t')\,k(t')\,f(t',t_b) = f_1(t_a,t_b),
\label{eq:fredholm}
\end{equation}
where $k$ and $f_1$ are assumed known and $f$ is the unknown. To solve 
for $f$ we discretize the time into $N$ total intervals of incremental 
length $\Delta t = t/N$. Applying the trapezoidal rule to the integral 
in Eq.~(\ref{eq:retnonequil}) we get
\begin{eqnarray}
\lefteqn{f(t_a,t_b) + \Delta t\cdot\left\{\frac{1}{2} f_1(t_a,t_0)\,k(t_0)
  \,f(t_0,t_b) \right.} \nonumber \\
  & & +~\sum_{j = 1}^{N-1} f_1(t_a,t_j)\,k(t_j)\,f(t_j,t_b)
  \nonumber \\
  & & \left.+~\frac{1}{2} f_1(t_a,t_N)\,k(t_N)\,f(t_N,t_b)\right\}
  = f_1(t_a,t_b),
\label{eq:trapezoid}
\end{eqnarray}
for a set of values of $t_a$ and $t_b$. We can recast the calculation 
into a linear problem of the form
\begin{widetext}
\begin{equation}
\left( \begin{array}{c}
  f(t_0,t_b) \\ f(t_1,t_b) \\ \vdots \\ f(t_N,t_b)
  \end{array} \right) + \Delta t\cdot\left( \begin{array}{cccc}
  \frac{1}{2} f_1(t_0,t_0)\,k(t_0) & f_1(t_0,t_1)\,k(t_1) &
  \cdots & \frac{1}{2} f_1(t_0,t_N)\,k(t_N) \\
  \frac{1}{2} f_1(t_1,t_0)\,k(t_0) & f_1(t_1,t_1)\,k(t_1) &
  \cdots & \frac{1}{2} f_1(t_1,t_N)\,k(t_N) \\
  \vdots \\
  \frac{1}{2} f_1(t_N,t_0)\,k(t_N) & f_1(t_N,t_1)\,k(t_1) &
  \cdots & \frac{1}{2} f_1(t_N,t_N)\,k(t_N)
  \end{array} \right)
  \left( \begin{array}{c}
  f(t_0,t_b) \\ f(t_1,t_b) \\ \vdots \\ f(t_N,t_b)
  \end{array} \right) = \left( \begin{array}{c}
  f_1(t_0,t_b) \\ f_1(t_1,t_b) \\ \vdots \\ f_1(t_N,t_b)
  \end{array} \right).
\end{equation}
\end{widetext}
We end up with a linear problem of the form
\begin{equation}
({\bf 1} + {\bf F})\cdot\vec{f} = \vec{f}_1.
\label{eq:linearprob}
\end{equation}
The unknown vector $\vec{f}$ can be determined by decomposing 
$({\bf 1} + {\bf F})$ using LU decomposition and then back substituting 
the result to the vector $\vec{f}_1$. The $G_{10}^{{\rm RL},r}(t_1,t_2)$ 
in Eq.~(\ref{eq:retnonequil}) is numerically determined this way for 
values of $t_1$ and $t_2$ within the interval $[0,t]$. The same 
calculation can also be done to determine the advanced Green's function 
$G_{10}^{{\rm RL},a}(t_1,t_2)$.

Having numerically calculated $G_{10,1}^{{\rm RL},<}$, 
$G_{10}^{{\rm RL},r}$, and $G_{10}^{{\rm RL},a}$, we can then use 
Eq.~(\ref{eq:green}) to determine $G_{10}^{{\rm RL},<}(t_1,t_2)$ and 
its derivative with respect to $t_2$. The current $I^{\rm L}(t)$ is 
then calculated from Eq.~(\ref{eq:leftcurrent}). The same steps can be 
followed to calculate the energy current flowing out of the right lead, 
$I^{\rm R}(t)$.

\section{Numerical results}
\label{sec:results}

We numerically calculate the time-dependent behavior of the energy
current flowing out of the leads. The coupling between the leads is
harmonically modulated in the form
\begin{equation}
k(t) = \frac{k}{2} (1 - \cos \omega_d t),
\label{eq:harmonic}
\end{equation}
where $\omega_d$ is the driving frequency. Note that this modulated 
coupling has the same order as $k$ and thus, a perturbative
calculation is not expected to produce accurate results. In contrast
to perturbative calculations, we calculate the current exactly by 
numerically solving the Dyson equation. In all of our calculations, 
we set the interparticle coupling $k = 0.625$ eV/({\AA}$^2$ u) and 
the on-site spring constant $k_0 = 0.0625$ eV/({\AA}$^2$ u). These 
choices lead to a natural time scale that we also use as the unit of 
time, $[t] = 10^{-14}~{\rm s}$. We use a time increment of 
$\Delta t = 0.1~[t]$ in our calculations. In addition, choosing the 
values of $k$ and $k_0$ also sets the width of the phonon band. In a 
linear chain, the phonon density of states is confined to be within 
$k_0 < \omega^2 < 4 k + k_0$.

\begin{figure}[h!]
\vspace{0.2in}
\includegraphics[width=3.3in]{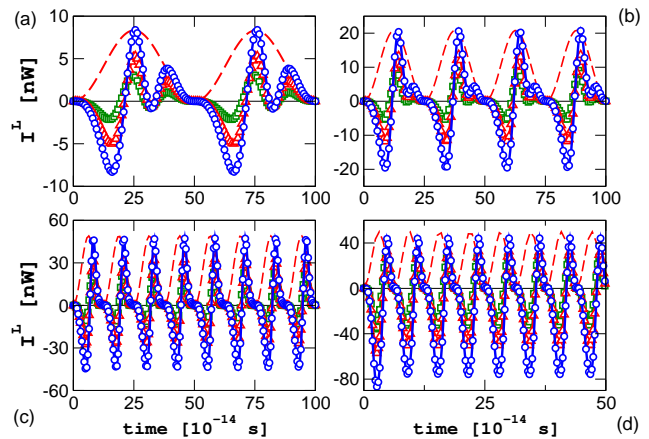}
\caption{(Color online) The current $I^{\rm L}(t)$ as a function of 
time when the driving frequency is (a) $\omega_d = 0.125~[1/t]$,
(b) $\omega_d = 0.25~[1/t]$, (c) $\omega_d = 0.5~[1/t]$, and
(d) $\omega_d = 1~[1/t]$. The leads have the same temperature
$T = T_{\rm L} = T_{\rm R}$ with values $T = 10$ K (green squares), 
$T = 300$ K (red triangles), and $T = 500$ K (blue circles). The 
harmonic modulation of the coupling is shown (red dashed line) as a 
guide. Its amplitude is not drawn to scale.
\label{fig:same_ncos_current}}
\end{figure}

We explore several variations of our setup. First, we study the energy
current when there is no thermal bias between the leads. Let the 
temperature of the leads be $T = T_{\rm L} = T_{\rm R}$. Shown in
Fig.~\ref{fig:same_ncos_current} is the time-dependent behavior of the 
current flowing out of the left lead for four different driving
frequencies and three different lead temperatures. Notice that the 
current does not exactly follow the modulation.

\begin{figure}[h!]
\vspace{0.25in}
\includegraphics[width=3.3in]{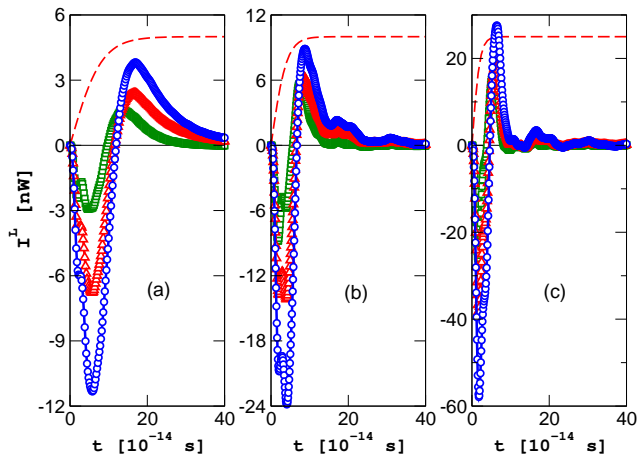}
\caption{(Color online) The current $I^{\rm L}(t)$ as a function of time 
$t$ when the coupling between the leads is gradually increased with
driving frequencies (a) $\omega_d = 0.125~[1/t]$, 
(b) $\omega_d = 0.25~[1/t]$, and (c) $\omega_d = 0.5~[1/t]$. Both leads
have the same temperature $T = 10$ K (green squares), $T = 300$ K (red 
triangles), and $T = 500$ K (blue circles). The gradually increasing 
coupling is shown (red dashed line) as a guide. Its amplitude is not 
drawn to scale.
\label{fig:same_tanh}}
\end{figure}

Although the amplitude of the modulated coupling is kept constant at
$k/2$, as the driving frequency is increased the peaks in the current 
also increases. In addition, as the driving frequency becomes 
sufficiently high, transient behavior in the current becomes visible.
Transient behavior becomes pronounced when the modulation of the
coupling changes rapidly. We look at transient behavior more closely
in Fig.~\ref{fig:same_tanh}.

\begin{figure}[h!]
\includegraphics[width=3.3in]{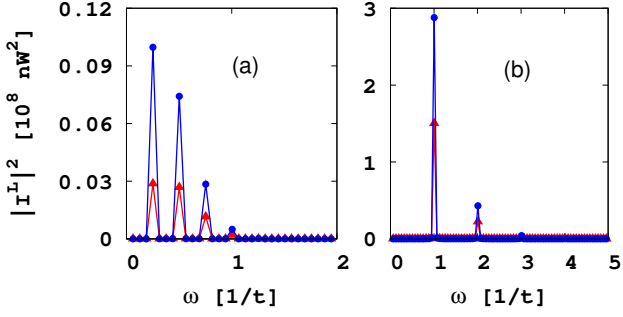}
\caption{(Color online) The square of the Fourier transform of the 
current $|I^{\rm L}[\omega]|^2$ as a function of the frequency 
$\omega$ when the driving frequencies are (a) $\omega_d = 0.25~[1/t]$ 
and (b) $\omega_d = 1~[1/t]$. The leads have the same temperature
$T$ and shown are when $T = 100$ K ($\blacktriangle$, red) and 
$T = 300$ K (\CIRCLE, blue).
\label{fig:same_spectrum}}
\end{figure}

Shown in Fig.~\ref{fig:same_spectrum} are plots of the Fourier
transforms of the left current when the driving frequencies are
$\omega_d = 0.25~[1/t]$ and $\omega_d = 1~[1/t]$. The peaks occur at
frequencies that are integer multiples of $\omega_d$. Modulating the
lead coupling therefore produces a dynamic energy current that is
composed of the first few harmonics of the driving frequency.

\begin{figure}[h!]
\vspace{0.2in}
\includegraphics[width=3.3in]{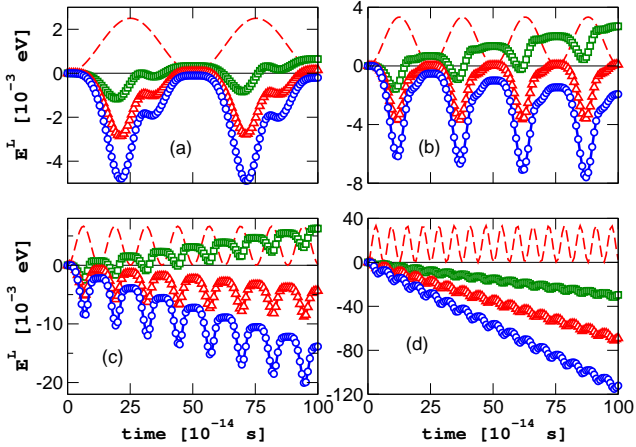}
\caption{(Color online) The energy $E^{\rm L}(t)$ as a function of 
time for driving frequencies (a) $\omega_d = 0.125~[1/t]$,
(b) $\omega_d = 0.25~[1/t]$, (c) $\omega_d = 0.5~[1/t]$, and
(d) $\omega_d = 1~[1/t]$. The leads have the same temperature $T$
and shown are for $T = 10$ K (green squares), $T = 300$ K (red 
triangles), and $T = 500$ K (blue circles). The harmonic modulation 
of the coupling is shown (red dashed line) as a guide. Its amplitude 
is not drawn to scale.
\label{fig:same_ncos_energy}}
\end{figure}

In Fig.~\ref{fig:same_ncos_current}, we notice that the current appears
to be more on the negative axis as the driving frequency is increased. 
Since we define the left current in Eq.~(\ref{eq:current1}) as the 
energy flowing out of the left lead, a negative value means that the 
energy is flowing into the lead. To be more definite, we calculate
how much energy has flowed into the left lead by
\begin{equation}
E^{\rm L}(t) = \int_0^t I^{\rm L}(t')\,dt'.
\label{eq:leftenergy}
\end{equation}
We again use the trapezoidal rule to numerically calculate the integral.
Shown in Fig.~\ref{fig:same_ncos_energy} is the energy $E^{\rm L}(t)$ for
the four different driving frequencies and three different temperatures
corresponding to those shown in Fig.~\ref{fig:same_ncos_current}. Notice
that for the highest driving frequency, as shown in 
Fig.~\ref{fig:same_ncos_energy}(d), the energy increases negatively in
time for all three temperatures shown. Therefore, harmonically modulating 
the lead coupling by a driving frequency of $\omega_d = 1~[1/t]$ moves 
energy from the external agent into the left lead. Furthermore, since 
there is no thermal bias between the leads, the left and right leads are
indistinguishable. The plots for $I^{\rm R}(t)$ and $E^{\rm R}(t)$ are 
therefore exactly the same as those shown for the left lead in
Fig.~\ref{fig:same_ncos_current} and Fig.~\ref{fig:same_ncos_energy},
respectively. The external agent therefore supplies energy to both the 
left and right leads.

Decreasing the driving frequency, we see from 
Fig.~\ref{fig:same_ncos_energy} that for certain values of the 
temperature, instead of the energy flowing into the leads, energy is 
actually flowing out from the leads. In 
Fig.~\ref{fig:same_ncos_energy}(c), for example, when the driving 
frequency is $\omega_d = 0.5~[1/t]$, energy flows into the leads when 
$T = 300$ K and $T = 500$ K but it flows out of the leads when $T = 10$ K.

\begin{figure}[h!]
\includegraphics[width=3.3in]{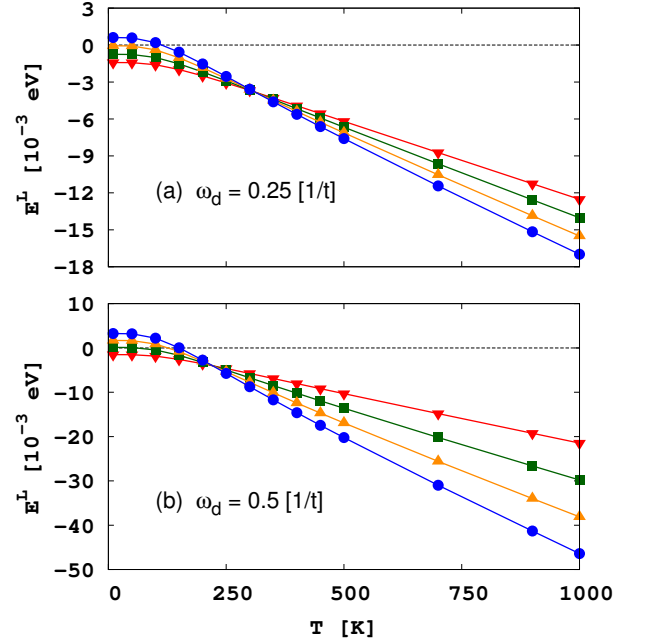}
\caption{(Color online) The energy $E^{\rm L}$ as a function of the 
lead temperature $T$, which is the same for both leads. Each curve in 
the plot corresponds to a specific value of time. (a) For 
$\omega = 0.25~[1/t]$ at time $t = 11.8~[t]$ ($\blacktriangledown$, 
red), $t = 36.9~[t]$ ($\blacksquare$, green), $t = 62.1~[t]$ 
($\blacktriangle$, orange), and $t = 87.2~[t]$ (\CIRCLE, blue). (b) For 
$\omega = 0.5~[1/t]$ at time $t = 19.4~[t]$ ($\blacktriangledown$, red), 
$t = 44.6~[t]$ ($\blacksquare$, green), $t = 69.7~[t]$ ($\blacktriangle$, 
orange), and $t = 94.8~[t]$ (\CIRCLE, blue).
\label{fig:same_tempdep}}
\end{figure}

Shown in Fig.~\ref{fig:same_tempdep} are plots of how $E^{\rm L}$ varies 
for different values of the driving frequency and lead temperature, at
specific instants of time. Note that in Fig.~\ref{fig:same_ncos_energy} 
the energy $E^{\rm L}$ oscillates in time. The data points in 
Fig.~\ref{fig:same_tempdep} are chosen from Fig.~\ref{fig:same_ncos_energy} 
during the times when $E^{\rm L}$ is at a trough in the oscillating 
energy. We can also choose different sets of data points, e.g., points 
at the crest instead of the trough, and then plot them like those in 
Fig.~\ref{fig:same_tempdep}. The plots, however, will look the same 
except for a translation along the vertical axis. 

In Fig.~\ref{fig:same_tempdep}, the $E^{\rm L}$ values are negative at 
higher temperatures. At lower temperatures, however, $E^{\rm L}$ can be 
positive, depending on the values of $\omega_d$ and $t$. Notice that
there is a temperature $T_c$ where the curves, for one value of 
$\omega_d$, intersect. In Fig.~\ref{fig:same_tempdep}(a), for example, 
$T_c \approx 300$ K. This $T_c$ is lower for higher $\omega_d$ values,
as shown in Fig.~\ref{fig:same_tempdep}(b). Now, since the choice of 
taking data points only at the trough of $E^{\rm L}$ is arbitrary, we may 
also choose a different location along $E^{\rm L}$ such that at $T_c$ we 
get $E^{\rm L}(T_c) = 0$, for any time $t$. For such a choice, the values 
of $E^{\rm L}$ below $T_c$ are positive and above $T_c$ are negative.

Notice in Fig.~\ref{fig:same_tempdep} that the slope of the curves
becomes steeper at later times. The value of $E^{\rm L}$ at $T = T_c$,
however, remains the same at any time $t$. Therefore, at much later 
times when the slope of the curves are much steeper, when $T > T_c$ both 
the left and right leads absorb energy from the external agent, while 
when $T < T_c$ the leads emit energy to the external agent.

\begin{figure}[h!]
\vspace{0.2in}
\includegraphics[width=3.3in]{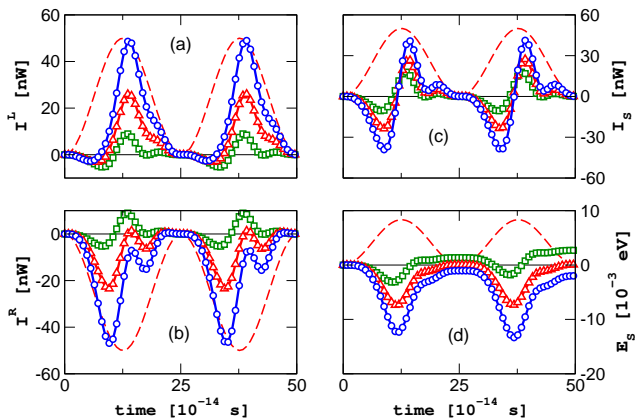}
\caption{(Color online) Time evolution of (a) the current in the left 
lead, $I^{\rm L}$, (b) the right lead, $I^{\rm R}$, (c) the sum of 
the currents, $I_s = I^{\rm L} + I^{\rm R}$, and (d) the sum of the 
energies in both leads, $E_s$. The average temperature between
the leads are $T_{\rm ave} = 10$ K (green squares), $T_{\rm ave} = 300$ K 
(red triangles), and $T_{\rm ave} = 500$ K (blue circles). The driving 
frequency $\omega_d = 0.25~[1/t]$. The harmonic modulation of the coupling 
is shown (red dashed line) as a guide. Its amplitude is not drawn to scale.
\label{fig:diff_ncos}}
\end{figure}

We now consider what happens when the leads have different temperatures.
Let the lead temperatures be
\begin{equation}
\begin{array}{c}
  T_{\rm L} = \left(1 + \varepsilon\right) T_{\rm ave}, \\
  T_{\rm R} = \left(1 - \varepsilon\right) T_{\rm ave},
\end{array}
\end{equation}
i.e., the left lead is warmer than the right lead. Heat would therefore
spontaneously flow, except for transient effects, from the left lead to 
the right lead if the coupling between the leads is not modulated in 
time \cite{cuansing10}. In our calculations we consider a temperature 
variation of $\varepsilon = 0.1$. Figure~\ref{fig:diff_ncos}(a) shows the 
current flowing out of the left lead as a function of time when the 
driving frequency $\omega_d = 0.25~[1/t]$. The time evolution of the 
current does not exactly follow the modulation of the lead coupling. 
Taking the Fourier transform of the current produces peaks at a few
integer multiples of the driving frequency. This situation 
is similar to the case when the leads have the same temperature, as
shown in Fig.~\ref{fig:same_spectrum}. Notice 
in Fig.~\ref{fig:diff_ncos}(a) that the current is mostly positive and 
thus, as expected, energy is flowing out of the warmer left lead. 
Shown in Fig.~\ref{fig:diff_ncos}(b) is the current flowing out of the 
right lead. For the cooler right lead, the question of whether the 
current is mostly negative or positive depends on the temperature of the 
lead. When $T_{\rm ave} = 10~{\rm K}$, the current is mostly positive 
and therefore, energy is mostly flowing out of the cooler right lead. At 
$T_{\rm ave} = 10~{\rm K}$ therefore, the current is flowing out of
both the left and right leads, i.e., energy from both leads is being
absorbed by the external agent. In contrast, when 
$T_{\rm ave} = 500~{\rm K}$ the current in the right lead is mostly 
negative and therefore energy is mostly flowing into the cooler right 
lead. 

Notice that the plots in Fig.~\ref{fig:diff_ncos}(b) are not mirror 
reflections, with respect to the horizontal axis, of the plots in 
Fig.~\ref{fig:diff_ncos}(a). When the sum of the currents are taken, i.e.,
$I_s(t) = I^{\rm L}(t) + I^{\rm R}(t)$, the results are the plots shown 
in Fig.~\ref{fig:diff_ncos}(c). Note that although the plots only show 
the time evolution of $I_s$ up to time $t = 50~[t]$ our calculations 
extend until time $t = 100~[t]$. Comparing Fig.~\ref{fig:diff_ncos}(c) to 
Fig.~\ref{fig:same_ncos_current}(b), we find the plots to coincide (note 
that the plot in Fig.~\ref{fig:same_ncos_current}(b) is only for the left 
lead and so the values of the current should be multiplied by two). 
Furthermore, we find that the values of $I_s(t)$ when $\varepsilon = 0.1$ 
are numerically very close to, if not the same as, the values of $I_s(t)$
when $\varepsilon = 0$ at each corresponding instant of time $t$. 
Our numerical results thus hint that the sum of currents $I_s(t)$ may be
independent of the value of $\varepsilon$.

Modulating the lead coupling when $\varepsilon = 0.1$ therefore
results in plots of the net current $I_s$ that numerically coincide
with those when $\varepsilon = 0$. The direction of the current flow in
each lead, however, differs depending on whether the leads have the same
or different temperatures. When the leads have the same temperature the
current can either flow into both of the leads at the same time or flow
out of the leads at the same time. In contrast, when the leads have
different temperatures, the energy can flow out of both the warmer left
lead and the cooler right lead resulting in a net current that flows 
out of the linear chain system and into the external agent. This happens,
for example, when $T_{\rm ave} = 10~{\rm K}$ in Fig.~\ref{fig:diff_ncos}. 
Increasing the average temperature to $T_{\rm ave} = 300~{\rm K}$, we find 
a balance between the energy that flows out of the warmer left lead and 
the energy that flows into the cooler right lead resulting in no net 
energy flow for the whole system, as shown in Fig.~\ref{fig:diff_ncos}(d). 
Increasing the average temperature further to $T_{\rm ave} = 500~{\rm K}$, 
we find that energy flows out of the left lead and flows into the cooler 
right lead resulting in a net energy flowing into the chain system. We
thus see that the current can either flow into or out of the cooler 
right lead depending on the value of the average temperature of the 
leads.

The driving frequency in Fig.~\ref{fig:diff_ncos} is 
$\omega_d = 0.25~[1/t]$. As shown in Fig.~\ref{fig:same_tempdep} there
is a temperature $T_c$ where there is no net energy flowing into or out
of the linear chain system. When $\omega_d$ is varied the $T_c$ also 
varies. Similarly, for the case when one of the leads is warmer than the
other, when $T_{\rm ave} < T_c$ we find that the current flows out of the 
cooler right lead resulting in a net energy flowing out of the chain 
system. When $T_{\rm ave} > T_c$ we find the current to flow into the 
cooler right lead resulting in a net energy flowing into the chain system.
Note that for any value of $T_{\rm ave}$ energy flows out of the warmer 
left lead.

We now investigate the effects of the speed of modulation on the 
transient behavior of the current. In Fig.~\ref{fig:same_ncos_current} 
we find that the transient becomes visible as the driving frequency 
$\omega_d$ is increased. To clearly see the effects of how fast the 
coupling is changing, we consider gradually increasing the coupling in 
the form
\begin{equation}
k(t) = k\,\tanh{\omega_d t}.
\label{eq:tanh}
\end{equation}
Shown in Fig.~\ref{fig:same_tanh} are plots of the energy current in
time for various values of the driving frequency $\omega_d$ and
temperature $T$. We consider the leads to have the same temperature 
$T$ and thus, for later times, we expect there to be no steady-state 
current. In Fig.~\ref{fig:same_tanh} we see that at earlier times the 
transient behavior shows up as rapid bumps in the current and then 
eventually subsides down to zero at later times. The amplitude of the 
transient current depends on the values of the driving frequency 
$\omega_d$ and the temperature $T$ of the leads. The faster $\omega_d$ 
and higher $T$ produce larger transient current amplitudes.

\vspace{0.1in}

\section{Summary}
\label{sec:summary}

Dynamically modulating the coupling between the leads in the 
form shown in Eq.~(\ref{eq:harmonic}) can result in the energy
current to either flow into or out of the leads, depending on the 
values of the driving frequency $\omega_d$ and the lead temperature 
$T$, even when the leads have the same temperature. For such a case,
it is possible for the energy current to either flow out of both leads 
at the same time or flow into both leads at the same time, as shown in 
Fig.~\ref{fig:same_ncos_energy}. In addition, in  
Fig.~\ref{fig:same_tempdep} we see that for a given value of $\omega_d$ 
there exists a temperature $T_c$ where, in the long-time limit, when the 
temperature of the leads $T < T_c$ we find the current to flow out of 
both leads and when $T > T_c$ we find the current to flow into both 
leads. For the case when the leads have different temperatures, with 
$\varepsilon = 0.1$, the direction of the flow of the energy current 
in the cooler lead depends on the values of $T_{\rm ave}$ and 
$\omega_d$. When $T_{\rm ave} < T_c$ the current flows out of the 
cooler lead but when $T_{\rm ave} > T_c$ the current flows into the 
cooler lead. Current flows out of the warmer lead for any temperature 
$T_{\rm ave}$. Gradually increasing the lead coupling in the form 
shown in Eq.~(\ref{eq:tanh}) shows that the amplitude of the transient 
depends on how fast the lead coupling changes. Faster changes in the 
lead coupling result in larger transient amplitudes, as shown in 
Fig.~\ref{fig:same_tanh}. As a consequence, harmonically modulating 
the lead coupling with a faster driving frequency results in a more 
pronounced transient behavior in the current, as shown in 
Fig.~\ref{fig:same_ncos_current}.

\begin{acknowledgments}

We would like to thank Jos\'{e} Garc\'{\i}a-Palacios, Lifa Zhang, 
Jin-Wu Jiang, Meng Lee Leek, Xiaoxi Ni, Bijay Agarwalla, and Juzar 
Thingna for insightful discussions. This work is supported in part by 
an NUS research grant number R-144-000-257-112. One of us (ECC) would 
like to thank the Department of Physics and Centre for Computational 
Science and Engineering at NUS where most of this work was done.

\end{acknowledgments}


\end{document}